\documentclass[useAMS,usenatbib]{mn2e}

\title[The spin-orbit angle of CoRoT-1]{The spin-orbit angle of the transiting hot jupiter CoRoT-1b\thanks{Based on observations obtained at the W.M. Keck Observatory and the European Southern Observatory}}
\author[F. Pont et al.]{F. Pont$^1$, M. Endl$^2$, W. D. Cochran$^2$, S. I. Barnes$^3$,  C. Sneden$^{4}$, P. J. MacQueen$^2$  \,   \newauthor
C. Moutou$^5$,  S. Aigrain$^1$, R. Alonso$^6$, A. Baglin$^7$,  F. Bouchy$^8$,  M. Deleuil$^5$, 
\newauthor   M. Fridlund$^9$,   G. H\'ebrard$^{10}$, A. Hatzes$^{11}$, T. Mazeh$^{12}$, A. Shporer$^{12}$\\
$^1$ School of Physics, University of Exeter, Exeter, EX4 4QL, UK \\
$^2$McDonald Observatory, The University of Texas at Austin, Austin, TX 78731, USA\\
$^3$The Anglo-Australian Observatory, 167 Vimiera Rd, Eastwood, NSW 2122, Australia\\
$^{4}$ Department of Astronomy, The University of Texas at Austin, Austin, TX 78731, USA\\
$^5$ Laboratoire d'Astrophysique de Marseille, UMR 6110, CNRS\&Univ. de Provence, 38 rue Fr\'ed\'eric Joliot-Curie, 13388 Marseille Cedex 13, France\\
$^6$ Observatoire de Gen\`eve, Universit\'e de Gen\`eve, 51 Chemin des Maillettes, 1290 Sauverny, Switzerland\\
$^7$ LESIA, CNRS UMR 8109, Observatoire de Paris, 5 place J. Janssen, 92195 Meudon, France\\
$^8$ Observatoire de Haute-Provence, 04870 Saint-Michel l'Observatoire, France\\
$^9$ Research and ScientiÞc Support Department, European Space  Agency, ESTEC, 2200 Noordwijk, The Netherlands \\
$^10$ Institut d'Astrophysique de Paris, UMR7095 CNRS, Universit\'e Pierre \& Marie Curie, 98bis boulevard Arago, 75014 Paris, France\\
$^{11}$ Th\"uringer Landessternwarte Tautenburg, Sternwarte 5, 07778 Tautenburg, Germany\\
$^{12}$ School of Physics and Astronomy, Tel Aviv University, Tel Aviv 69978, Israel}

\def\m2s2{\hbox{\,m$^{2}$\,s$^{-2}$}} %m2.s -2
\def\kms{\hbox{\,km\,s$^{-1}$}}       %km.s -1
\usepackage{graphicx}
\usepackage{txfonts}
\usepackage{natbib}

\usepackage{epsf}
\usepackage{rotating}
\usepackage{graphics}
\begin{document}

\date{}

%\pagerange{\pageref{firstpage}--\pageref{lastpage}} \pubyear{}

\maketitle

\begin{abstract}
We measure the angle between the planetary orbit and the stellar rotation axis in the transiting planetary system CoRoT-1, with new HIRES/Keck and  FORS/VLT high-accuracy photometry. The data indicate a highly tilted system, with a projected spin-orbit angle $\lambda = 77 \pm 11 ^\circ$.  Systematic uncertainties in the radial velocity data could cause the actual errors to be larger by an unknown amount, and this result needs to be confirmed with further high-accuracy spectroscopic transit measurements.
 Spin-orbit alignment  has now been measured in a dozen extra-solar planetary systems, and several  show strong misalignment. The first three misaligned planets were all much more massive than Jupiter and followed eccentric orbits. CoRoT-1, however, is a jovian-mass close-in planet on a circular orbit.  If its strong misalignment is confirmed, it would break this pattern. The high occurence of misaligned systems for several types of planets and orbits favours planet-planet scattering as a mechanism to bring gas giants on very close orbits.  %Our extensive radial-velocity monitoring of CoRoT-1 excludes the presence of another gas giant planet in the system out to about~2~AU.
\end{abstract}


\section{Introduction}

The projection on the plane of the sky of the angle between a planetary orbit and the rotation axis of its host star can be measured in transiting systems, using the Rossiter-McLaughlin (``RM'') effect  \citep{ross24,mcla24}. The passage of the planet in front of the star produces an anomaly in the radial velocity curve that depends on this angle. The distribution of spin-orbit angles is an important clue to the formation and evolution mechanisms of planetary systems. Close-in gas giant planets are thought to be formed in a disc and migrate inwards by interaction with the disc \citep{lin96}, a scenario  expected to produce orbits with generally aligned spin  and orbit. 

The projected spin-orbit angle has now been measured precisely for more than a dozen extra-solar planetary systems. Starting with XO-3 \citep{hebr08}, a substantial fraction of these systems exhibit large spin-orbit misalignments \citep{pont09,john09}. The most extreme case to date is the retrograde orbit of HAT-P-7b \citep{nari09, winn09b}.

CoRot-1b\footnote{Originally known as CoRoT-Exo-1 b. The naming convention for CoRoT planets has been subsequently modified.} \citep{barg08} is the first transiting planet identified by the CoRoT space mission.  It is an extreme representative of the ``hot Jupiter'' family of extra-solar planets, with an orbital period of only 1.5 days. Its mass is similar to that of Jupiter but, like other  gas giants on very close orbits, it has a much larger radius ($R\sim 1.4$ R$_J$). At $V=13.6$ mag, the parent star lies at the faint end of the magnitude range for presently known planet hosts, so that follow-up observations require large telescopes. As part of the ground-based follow-up of the CoRoT planet search, new photometric and spectroscopic observations have been gathered on this target, with the objective of refining the system parameters, notably the planet size, of detecting any further longer-period planet in the system, and of measuring the spin-orbit angle through the RM anomaly. 
We have monitored one transit with the FORS camera on the VLT in two passbands and measured the spectroscopic transit with the HIRES spectrograph on the 10-m Keck telescope. % In addition to the radial velocity data published in  \citet{barg08}, acquired with the SOPHIE (OHP, France) and HARPS (ESO, Chile) radial-velocity spectrographs, CoRoT-1 has also been monitoring with the HARPS spectrograph as part of  Guaranteed Time Observations (M. Mayor, priv. comm.).
An analysis of the photometric transit curve to constrain the planet size was already presented by \citet{gill09}. In this study we concentrate on the determination of the spin-orbit angle from the HIRES data,  using the FORS data to contrain the other transit parameters. %Since the methodology to analyze this type of data has become well established by numerous previous observational studies of transiting extra-solar planets, we do not dwell on the technical details and refer the reader to the literature.

\section{Observations}

\subsection{Photometry}

%\subsubsection{FORS data}

The transit was measured in high-accuracy photometry with the FORS camera on the VLT (ESO, Chile) in the $R$ and $B$ filters on February 27, 2008. The observing strategy and technical details are identical to those described in \citet{pont07}. The reduction was described in  \citet{gill09}, and the data is given in Table~\ref{phot_tab}. We chose to alternate between the $R$ and $B$ filter because the difference in transit lightcurve shape between red and blue wavelengths, caused by the wavelength dependence of the  amount of stellar limb darkening, offers a useful complementary constraint on the transit impact parameter, in complement to the shape of the transit ingress and egress.  The mean time interval between successive measurements is 78 seconds. The dispersion of the residuals is 0.56 mmag in R and 0.52 mmag in B, close to the photon-noise limit.

%\begin{figure}
%\resizebox{8cm}{!}{\includegraphics{corot1_vr.pdf}}
%\caption{Radial velocity data from the HARPS (circles) and HIRES (squares) spectrographs. The uncertainties shown do not take into account the systematic errors, which dominate the final error budget (they are accounted for in the model fit). The line traces the best-fit radial-velocity orbit, including the RM anomaly.}
%\label{rv}
%\end{figure}

\begin{figure}
\resizebox{9cm}{!}{\includegraphics{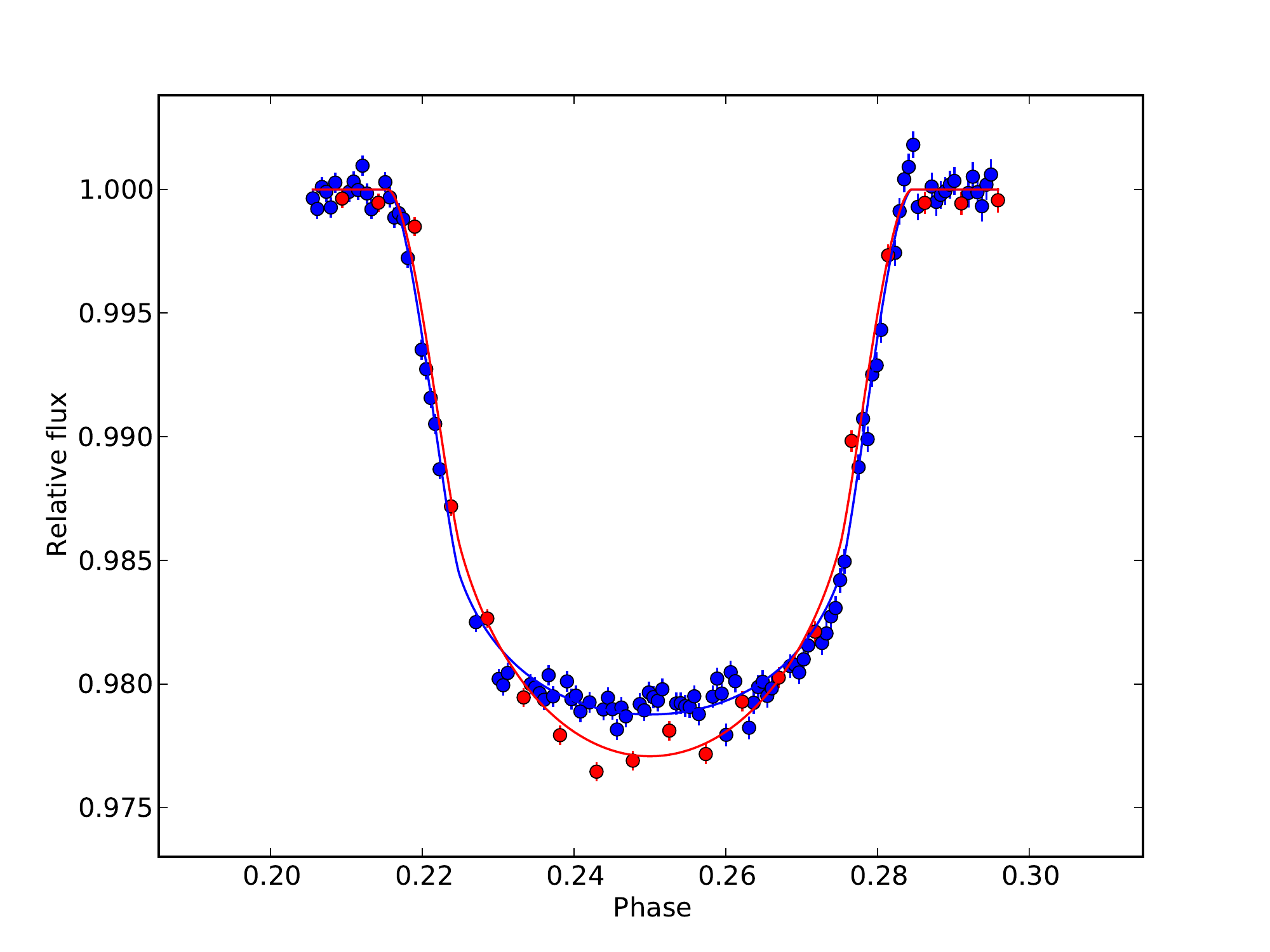}}
\caption{Photometric data from FORS/VLT in the B (squares) and R (circles) filters, with the best-fit transit models. }
\label{phot}
\end{figure}

\begin{figure}
\resizebox{9cm}{!}{\includegraphics{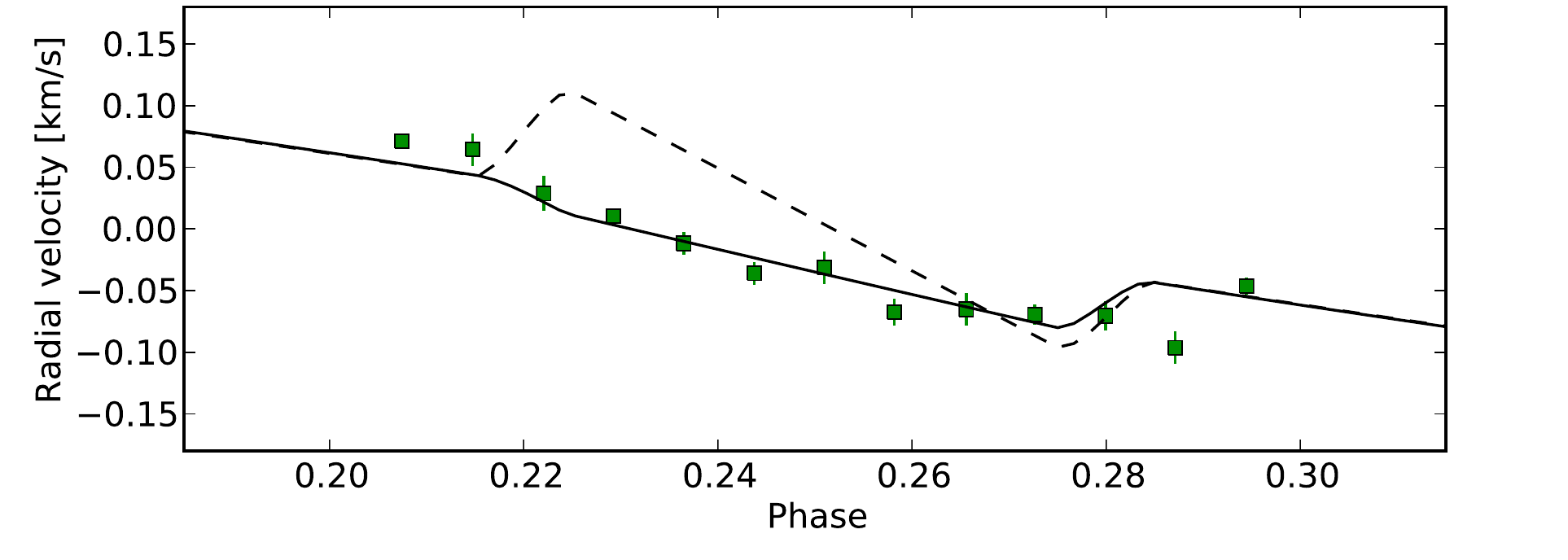}}
\caption{Radial velocity data around the phase of transit. The solid line shows the best-fitting models. The dashed line shows the best fit with $\lambda=0$ and $V_{\rm rot} \sin I_s =5.2$ km$\,$s$^{-1}$ imposed. }
\label{rm}
\end{figure}

\begin{table}
\centering
\begin{tabular}{lll} 
\hline
Date [HJD] & RV [m$\,$s$^{-1}$]& $\sigma_{RV}$ [m$\,$s$^{-1}$]  \\ \hline 
2454839.930643 & $-$2448.81 & 17.94\\
... & ... & . ... \\ \hline
\end{tabular}
\caption{Radial velocity measurements from HIRES/Keck and HARPS  (full table available electronically). }
\label{rv_tab}
\end{table} 

\begin{table}
\centering
\begin{tabular}{llll} 
\hline
Date [HJD] & Relative flux & uncertainty & Filter \\ \hline 
2454524.556083 & 0.9996 & 0.0004 & $R$ \\
%2454524.556973 & 0.9992 & 0.0004 & $R$ \\
%2454524.557873 & 1.0001 & 0.0004  & $R$\\
%2454524.558763 & 0.9999 & 0.0004  & $R$\\
%2454524.559653 & 0.9992 & 0.0004  & $R$\\
... & ...& ...  & ... \\ \hline
\end{tabular}
\caption{Photometric data from FORS in $B$ and $R$ (full table available electronically).}
\label{phot_tab}
\end{table}

\subsection{Radial velocity}

The radial velocity data comes from  the initial follow-up spectroscopy of CoRoT planet candidates with the SOPHIE and HARPS spectrographs, presented in \citet{barg08}, and a measurement of the spectroscopic transit with the Keck~1 telescope. In addition, we used radial velocity measurements obtained in 2008 with HARPS  (ESO Prog. 082C-0.312) to constrain the radial-velocity orbit outside the time of transit.
%\subsubsection{HARPS data}
%HARPS is a high-accuracy radial velocity spectrograph installed on the ESO 3.6m telescope at La Silla, Chile. CoRoT-1 was observed twenty-four times in the 2007/8 and 2008/9 seasons.
%HARPS can reach a precision of 1~\ms\ or better for bright stars. For stars as faint as CoRoT-1, however, background light and detector effects become a significant source of error, and systematics at the 50 \ms\ level are present in the data, even after correction for background light and other effects that we were able to identify. Such systematics are small compared to the scale of the orbital signal.

On 2009 January 8, we used Keck 1 and its HIRES spectrometer to observe the RM effect of CoRoT-1. % In the magnitude range relevant for CoRoT-1, this combination allows higher signal to noise ratios and lower systematics than HARPS. 
We used the same instrument setup to measure precise RVs as the California Planet Search group (G. Marcy. priv. comm.). An iodine gas absorption cell is placed in the optical path to provide a precise wavelength calibration. We obtained 13 spectra during the transit with an exposure time of 900 seconds, a slit width of 0.86 arcsec yielding a resolving power of R$\sim$45000, and a typical S/N ratio of $\approx 40:1$ in the iodine region. Precise differential RVs were computed with our {\it Austral} iodine code \citep{endl00}. Table~\ref{rv_tab} gives the HIRES/Keck and new HARPS radial velocity data. %The HARPS and SOPHIE data are available on request to the authors and will be published elsewhere as part of the CoRoT ground-based follow-up.

\section{Analysis}

The procedure to infer physical parameters from observations of transiting planetary systems has now become standard, and detailed descriptions can be found in the recent literature on the topic \citep[see ][ for a recent review]{won09}. A Monte-Carlo Markov Chain (MCMC) algorithm is used to integrate the posterior probability distribution for the model parameters given the observed data and other constraints. Correlated systematics in the data are accounted for by adapting the merit function to include the effect of red noise. Our implementation is described in more details in \citet{pont09} and only briefly summarized here. We use the \citet{mandel} and \citet{ohta} formalisms for the transit and RM effect model curves. The limb-darkening parameters are fixed to the values in \citet{claret}. %($u_{1,2}= 0.40, 0.27$ in $R$, $u_{1,2}=0.72, 0.10$ in $B$)
 Stellar evolution models \citep{gir00} are used to constrain the  host star's mass and radius. We add a factor in the merit function to take the correlation of the noise into account, using a set of $\sigma_r$ parameters \citep{pont06}, one for each instrument. We use the Metropolis-Hasting jump rejection criterion for the MCMC chain, and adopt a chain length of $10^5$. We verified that the correlation length for all fitted parameters is much shorter than the scale length.

Based on the residuals around the model curves, we find that systematics dominate photon-noise uncertainties in the radial velocity data. As discussed in \citet{pont06}, it is important to take the systematics and their time correlation into account to obtain reliable parameters and uncertainties. The effect of the systematics depends mainly of their correlation in time. There are not enough measurements in the radial-velocity transit sequences to measure the correlation length from the data itself. At the two extremes, we can either assume that the systematics are dominated by timescales very different from the transit duration, or that the systematics are dominated by timescale comparable to the transit duration. If, following \citet{winn08}, we note $\beta$ the ratio between the effective uncertainties to be used in the fit to the photon-noise uncertainties, in the first case we have $\beta \simeq {\sigma_{\rm res}}/{\sigma_{\rm pn}}$, in the second  $\beta \simeq n^{1/2} {\sigma_{\rm res}}/{\sigma_{\rm pn}}$, where $ \sigma_{\rm pn}$ is the photon-noise uncertainty on single data points, $\sigma_{\rm res} $ the r.m.s. of the residuals, and $n$ the number of measurements during the transit. For our standard solution, we use the first value (negligible time correlation of the radial velocity systematics).  In our experience, systematic errors for radial velocity measurements are less strongly correlated in time than photometric errors. We also examine the effect of a higher correlation.

We fit for the following system parameters: four orbital parameters: $T_{\rm tr}$ (epoch of transit), $V_0$ (centre-of-mass velocity), $K$ (velocity semi-amplitude), $i$ (orbital inclination);
mass and radius of the star $M_s, R_s$;
 mass and radius of the planet $M_p, R_p$;
and projected spin-orbit angle $\lambda$.
 We set the orbital period to the value found from CoroT photometry, $P=1.5089557$ days, and the orbital eccentricity to zero. The best constraint on eccentricity comes from the timing of the secondary eclipse, which indicates a negligible eccentricity \citep[$e \cos \omega< 0.01$,][]{gill09}.  We use a flat prior in age and metallicity for CoRoT-1, as well as the temperature and stellar rotation velocity  measured in \citet{barg08}: $T_{\rm eff}=5950 \pm 50$~K, $V_{\rm rot} \sin I_s = 5.2 \pm 1.0$ \kms. %These constraints are added as additional terms in the prior probability density: $P(V sin I;T) \sim \exp\{-\frac{(V sin I;T_{obs}- V sin I)^2}{2\sigma^2_{V \sin ;T}}\}$.    

\section{Results and discussion}

\subsection{System parameters}

%Table~\ref{param} gives the best value (best-fit value for the orbital parameters, median of the posterior probability function for the physical parameters of the star and planet) and central 68\% confidence intervals for the parameters of the CoRoT-Exo-1 system given by the MCMC integration, in the case assuming uncorrelated radial-velocity systematics. 

%
%\begin{table}
%\centering
%\begin{tabular}{l  l l }
%\hline
%Center-of-mass velocity $V_r$ (Sophie) & 23.592 $\pm$ 0.022 km$\,$s$^{-1}$ \\ 
%$V_0$(HARPS)& 23.506 $\pm$ 0.008 km$\,$s$^{-1}$\\ 
%Orbital period $P$ &1.5089557 (fixed) \\
%Orbital eccentricity $e$ &0 (fixed)  \\
%Velocity semi-amplitude $K$ &198$\pm$ 9 m$\,$s$^{-1}$\\ \\
%Orbital inclination $i$  & 86.7 $\pm$ 0.6 $^\circ$\\
%Semi-major axis $a$ & 0.0261 $\pm$ 0.0005 AU \\
%Epoch of transit & 2454524.2459 $\pm$ 0.0002 BJD \\ 
%Radius ratio $R_p /R_s$ &0.134 $\pm$ 0.001 \\ 
%{\bf Spin-orbit alignment $\lambda$ }& 77 $\pm$ 11 $^o$  \\
%Impact parameter $b$ & 0.35 $\pm$ 0.08 \\
%Star Mass $M_s$& 1.03 $\pm$ 0.06 M$_\odot$ \\
%Star Radius $R_s$ &1.14 $\pm$ 0.03 R$_\odot$ \\
%Planet mass $M_p$ &1.13 $\pm$ 0.07 $M_J$ \\
%Planet radius $R_p$  & 1.48 $\pm$ 0.06 $R_J$ \\ \hline
%\end{tabular}
%\caption{Parameters for the CoRoT-Exo-1 system. Uncertainties from the 68\% central probability interval of the posterior distribution described by the Markov chain. The uncertainties include the effect of uncorrelated systematics.}
%\label{param}
%\end{table}

The best values and central 68\% confidence intervals for the parameters of the CoRoT-1 system given by the MCMC integration, in the case assuming uncorrelated radial-velocity systematics, are the following:  
orbital elements $K$ =198 $ \pm $ 9 m$\,$s$^{-1}$,
 $i = 86.7 \pm0.6^\circ$,
 $a = 0.0261 \pm$ 0.0005 AU, 
$T_{\rm tr} = 2454524.6231\pm$ 0.0002 BJD, impact parameter $b =0.35 \pm$ 0.08,
star's mass and radius $M_s$ = 1.03 $\pm$ 0.06 M$_\odot$,
$R_s$ =1.14 $\pm$ 0.03 R$_\odot$,
planet's mass and radius $M_p =1.13 \pm$ 0.07~M$_J$,
$R_p = 1.48 \pm 0.06 $R$_J$.
We find $\lambda$  = 77 $\pm 11^\circ$ for the projected spin-orbit angle. Figures \ref{phot} and \ref{rm} compare the data and best-fit model. Figure~\ref{rm} also shows, for comparison, the best-fit model of the spectroscopic transit with an aligned orbit.%, and the best-fit with a free zero-point shift for the night of the HARPS spectroscopic transit measurements (which correspond to allowing for a long correlation timescale in the systematics).
%We repeated the integration allowing for a secular drift in the radial velocity zero-point, due to the orbit of a second, longer-period planet. We find no evidence for secular drift, to the level of 20~\ms\ over two years. The residuals around the best fit as a function of date are shown in Fig.~\ref{second}.

\begin{figure}
\resizebox{8cm}{!}{\includegraphics{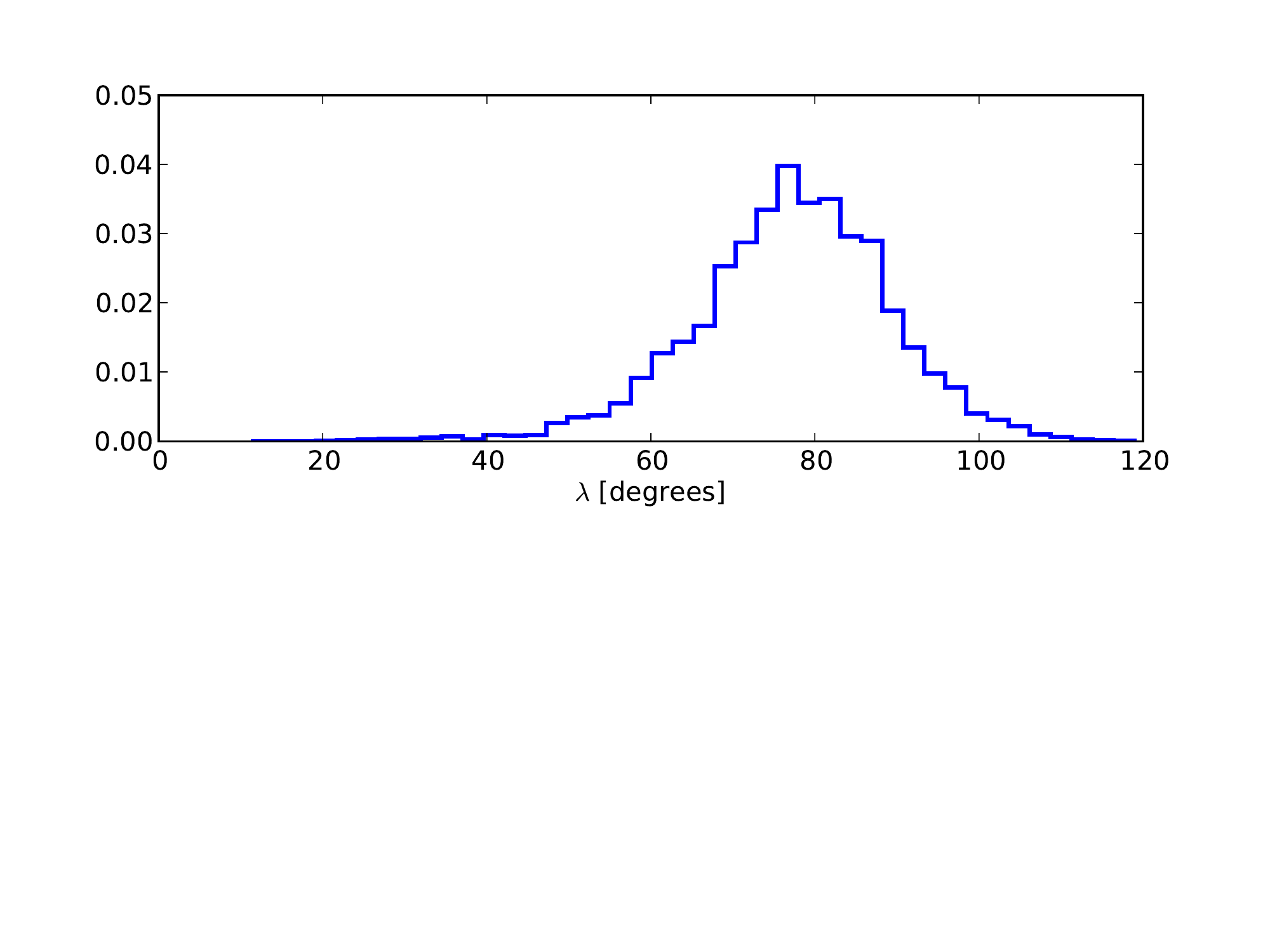}}
\vspace{-2.8cm}
\caption{Probability distribution of the projected spin-orbit angle according to the MCMC integration. }
\label{lam}
\end{figure}

%\begin{figure}
%\resizebox{8cm}{!}{\includegraphics{maxmass.pdf}}
%\caption{Posterior probability distribution:  drift}
%\label{drift}
%\end{figure}

%\begin{figure}
%\resizebox{8cm}{!}{\includegraphics{corot1_RMfit.pdf}}
%\caption{Radial velocity data around the time of transit with best-fit model ($\lambda=85^\circ$). Symbols as in Fig.~\ref{rv}. }
%\label{RM}
%\end{figure}

%

%\begin{figure}
%\resizebox{6cm}{!}{\includegraphics{config.pdf}}
%\caption{System configuration according to the best-fit model ($R_p/R_s=0.13, b=0.23, \lambda=78^\circ$).}
%\label{config}
%\end{figure}

The results on the mass and radius of the star and planet are comparable to previous determinations. Since we use the same values of the spectroscopic parameters of the host star as \citet{barg08}, a significant update will depend on an improved determination of the spectroscopic parameters. 

We discuss below the contraint on the spin-orbit angle of the system.

\subsection{Spin-orbit misalignment}

Figures~\ref{lam} shows the posterior probability distribution function for the projected spin-orbit angle. Our MCMC integration on all available data and constraints favour an angle peaking at $\lambda=+77^\circ$ with a probability distribution well described by a Gaussian with $\sigma \simeq 11^\circ$. In this solution, the star is tilted sideways relative to the planetary orbit, with the planet crossing only the receding side of the star. 

Two factors have an important influence on the values derived for the spin-orbit angle: the measured stellar rotation velocity, and the radial velocity measurement systematics during the transit. 

\subsubsection{ Rotation velocity of CoRoT-1}

A rotation velocity of $V_{\rm rot} \sin I_s =5.2 \pm 1.0$ km$\,$s$^{-1}$ is measured from the broadening of the HARPS spectral cross-correlation function (where $I_s$ is the angle between the line-of-sight and the rotation axis of the star, the third angle specifying the spin-orbit configuration, together with $\lambda$ and planet orbital angle $i$).  Since this constraint has an important influence on the output value for the spin-orbit angle, it is worth examining more closely. A lower value of the rotation would make the relatively flat velocity curve during the transit compatible with lower values for the spin-orbit angle. The measurement of stellar rotation from the broadening of the cross-correlation function has a long pedigree in radial-velocity studies, and the calibration for HARPS is well established. An analysis of the HIRES spectra also indicate significant rotation, $  V_{\rm rot} \sin I_s \simeq 6 \pm 1$~km$\,$s$^{-1}$.  We therefore consider the determination of the star's projected rotation velocity to be robust.%We also re-examined the CoRoT lightcurve in search for some photometric indication of the rotation period through spot-induced variations. %Using the new cleaning algorithm of \citet{alap09} for CoRoT lightcurve, 
%We find a possible signal near 10 days in the Lomb-Scargle periodogram, which would correspond to $V_{\rm rot} \sim$ 5 km$\,$s$^{-1}$, compatible with the spectroscopically measured value for $\sin I_s \sim 1$. 
%Figure~\ref{aude} shows the power spectrum of the detrended CoRoT lightcurve for CoRoT-1.

%\begin{figure}
%\resizebox{8cm}{!}{\includegraphics{blank.pdf}}
%\caption{Power spectrum for the CoRoT light curve of CoRoT-1, processed with the Alapini et al. algorithm.}
%\label{aude}
%\end{figure}

%The relatively flat Keck velocity curve could look suspicious with a rotation velocity of several km$\,$s$^{-1}$, since it implies some compensation between the rotation velocity and the spin-orbit angle. It could argue in favour of a smaller rotation velocity. Nevertheless, the Keck time series is not a straight line, and the departure is significant. 

If the  a priori constraint on $V_{\rm rot} \sin i$ is relaxed, the MCMC integration still converges towards highly tilted spin-orbit angles, with lower values for the rotation velocity in the 2-3 \kms\ range. Therefore, the spectroscopic data, especially the Keck velocity curve, favour a high spin-orbit tilt even without  external constraint on the stellar rotation velocity. %Figure~\ref{vsini0} shows the posterior probability distribution from a MC chain without any a priori constraint on $V sin I_S$.

%\begin{figure}
%\resizebox{8cm}{!}{\includegraphics{vsini_vs_lambda.pdf}}
%\caption{Posterior probability distribution for the projected rotation velocity and spin-orbit angle, without $V\sin I_s= 5.2 \pm 1 $km$\,$s$^{-1}$ used as an a priori constraint.}
%\label{vsini0}
%\end{figure}

%Therefore, empirical indications converge towards the reality of the measured value of  $V_{\rm rot} \sin i$ near 5 km$\,$s$^{-1}$, and the robustness of the spin-orbit angle result. 

\subsubsection{Red noise in the radial velocity data}

As discussed in the case of HD 80606 in \citet{pont09}, the posterior probability distribution for the  spin-orbit angle depends on the choice of the assumptions on the amplitude and timescale of the time-correlated  noise in radial velocity during the transit.

 If $\sigma_r$ is of the same order as the expected RM effect ($\sim$ 50 m$\,$s$^{-1}$), and correlated in time over similar timescales, then obviously the constraint on the spin-orbit angle becomes unreliable.% (this is of course just another way of saying that an effect cannot be reliably measured if unknown systematics are just as large). 

%The HARPS data during transit is relatively noisy, and systematics of up to 50 m$\,$s$^{-1}$\ or higher seem present. This is due to the target being near the faint end of the HARPS regime. 

%The Keck data are subject to smaller systematics.
Since the time series is not much longer than the duration of the transit, it is not possible to determine the correlation timescale on the data itself, but the point-to-point scatter of the HIRES data seems to indicate correlated systematics on the 10-20 m$\,$s$^{-1}$ scale at most. This is similar to the scatter that we found on other targets from the CoRoT follow-up observed with the same instrument.

We have repeated the MCMC integration using a correlation timescale comparable to the duration of the transit. Because the posterior probability distributions are relatively well-behaved and unimodal (see for instance Fig.~\ref{lam} for the spin-orbit angle), the effect is roughly equivalent to multiplying the parameter uncertainties by a factor $n^{1/2}$. With extreme assumptions on the correlation of the noise ($n = 10$), the result of a spin-orbit misalignment is still significant, but at a reduced level of $\sim 2 \sigma$.

%The agreement between the HARPS and HIRES data sets on the overall shape of the spectroscopic transit - notably 
%The absence of detectable velocity changes after  during ingress or egress  lends additional credence to the fact that correlated systematics do not dominate the measurement of the spin-orbit tilt. 

%\begin{figure}
%\resizebox{8cm}{!}{\includegraphics{corot1_RM0.pdf}}
%\caption{Radial velocity data around the time of transit compared to a model with aligned spin and orbit ($\lambda=0$) and $V\sin I_s= 5.2 $km$\,$s$^{-1}$. }
%\label{RM0}
%\end{figure}
Figure ~\ref{rm} compares the radial velocity data with the expected behaviour in case of an aligned orbit, assuming $V_{\rm rot} \sin I_s = 5.2$ km$\,$s$^{-1}$. Explaining the data with such a model would require not only high systematics, but a remarkable amount of compensation, that we consider unlikely.  However, the four data points outside the transit do now follow the model closely, which indicates that we do not have a tight control over the systematics at the point. 

  One important argument in favour of a strong misalignment is the result from the HARPS spectroscopic transit \citep{bou09}. Although the HARPS data is more noisy, it also shows a radial velocity constantly {\it lower} than the Keplerian orbit during the transit, indicating a planet crossing only the receding half of the star.

Precise and high-cadence radial velocity measurements for an object fainter than 13th magnitude in the visible are difficult even with the best instruments and largest telescopes. As for previous cases of measured spin-orbit misalignment, XO-3  \citep{hebr08,winn09}, HD 80606 \citep{mout09,winn092} and WASP-14 \citep{john09}, further observations and an independent confirmation of our result will be useful.

\begin{table}
\begin{tabular}{lrr}
\hline
System & $\lambda [^\circ]$ & $e$ \\ \hline
{\it circular and aligned} & & \\
HD 189733b& $-$1.4 $\pm$ 1.1  &$<$ 0.03\\
HD 209458b &0.1 $\pm$ 2.4 &$<0.03$\\
HAT-P-1b &3.7 $\pm$ 2.1  & $<0.10$\\
CoRoT-Exo-2b &7.2 $\pm$ 4.5 &$<0.10$ \\
HD 149026b &1.9 $\pm$ 6.1 &$<0.18$ \\
TrES-2b &$-$9.0 $\pm$ 12.0  &$<0.08$ \\
%TrES-1b &30.0 $\pm$ 21.0 & $<0.08$ \\
{\it circular and tilted} & & \\
{\bf CoRoT-1b} &77 $\pm$ 11 &$<0.01$ \\
{\bf HAT-P-7b}& 182.5  $\pm$ 9.4 &  $\sim 0 $\\
% & & \\
 {\it eccentric and aligned} & & \\
HAT-P-2b &1.2 $\pm$ 13.4 & 0.52  \\
HD 17156b &9.4 $\pm$ 9.3& 0.67\\
 {\it eccentric and tilted} & & \\
{\bf WASP-14b} & $-$33 $\pm$ 7 & 0.09 \\ 
{\bf XO-3b} &37.3 $\pm$ 3.7  & 0.23 \\
{\bf HD 80606} & 39 $^{+28}_{-13}$ & 0.93 \\ \hline
\end{tabular}
\caption{Transiting exoplanets with significant projected spin-orbit angle determinations (formal errors smaller than $20^\circ$). Angles from the compilation in \citet{fabr09}, updated with \citet{john09,winn092} and this paper. Eccentricities from \citet{madh09},  except \citet{josh09} for WASP-14. }
\label{meas}
\end{table}

\subsection{Discussion}

The projected spin-orbit angle has now been measured with adequate accuracy for more than a dozen transiting planets. %The current robust determinations are listed in Table~\ref{meas} (we do not include still highly uncertainties estimates, such as TrES-1, CoRoT-3, WASP-17, since these would introduce a bias towards misalignment).
%\footnote{We have re-analysed the cases of HD 189733, and CoRoT-3, presented in \citet{tria09}. These authors do not take red noise into account in their analysis, and therefore highly underestimate their uncertainties. They quote a projected spin-orbit angle of $VALUE$ for HD 189733. As visible from their figure NUM however, the instrumental red noise is of the same amplitude as the difference between the best-fit model velocity curve and a curve corresponding to spin-orbit mislignment. Re-doing the calculate with an account for red noise would give a more realistic uncertainty estimate, but common sense alone should be sufficient to tell that, if the systematics are similar to the measured effect, then the effect is not very significant. For CoRoT-3, the already hardly significant value of the misalignment is made even less significant by accounting for red noise.}.
The main result is the strikingly frequent occurrence of high spin-orbit misalignments. At least four systems have spin-orbit tilts in excess of 30 degrees (XO-3, HD 80606, WASP-14, HAT-P-7, see Introduction). Evidence for misalignment may also have been found for WASP-17 \citep{hell09}. 
According to the data and analysis presented here, CoRoT-1 joins this group, with the radial velocity data favouring an almost polar orbit. 

\begin{figure}
\resizebox{8cm}{!}{\includegraphics{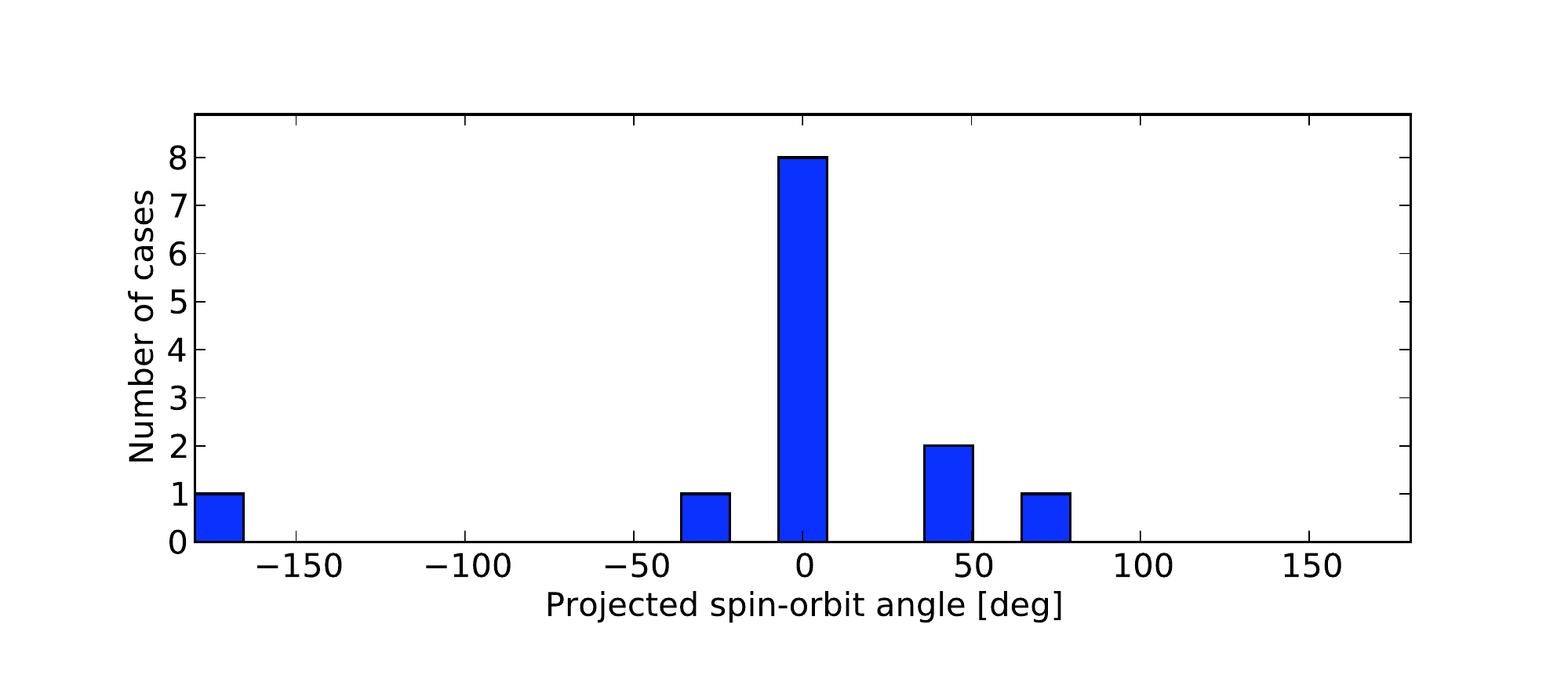}}
%\vspace{-2.8cm}
\caption{Distribution of projected spin-orbit angles for systems in Table~\ref{meas}.}
\label{hist_lam}
\end{figure}

Table~\ref{meas} displays the systems with significant spin-orbit angle measurements, in relation with orbital eccentricities. Figure~\ref{hist_lam} shows this list in histogram form. Note that it is important for this ensemble analysis to include only conclusive determinations that do account for correlated noise in the uncertainties.  Values with underestimated uncertainties introduce an obvious bias towards non-alignment.

\citet{john09} pointed out an interesting feature of the first three misaligned systems (XO-3, WASP-14 and HD 80606): all contained a very high-mass gas giants (4, 7 and 12 Jupiter masses) on markedly eccentric orbits. This suggested that misalignment was not a feature of ordinary hot jupiters, but of their higher-mass cousins.
CoRoT-1, however,  would break this pattern. It has a mass nearly equal to Jupiter, and a circular orbit. The same is true for  HAT-P-7 \citep{nari09,winn09b}. Therefore the correlation between planet mass, eccentricity and misalignment  is called into question.

The cases of CoRoT-1 and HAT-P-7 reinforce the statistical conclusions drawn on an earlier sample by \citet{fabr09}: the sample as a whole is compatible with a bimodal distribution of spin-orbit angles, with some systems well aligned, and others with little correlation between the two angles. The distribution illustrated by Fig.~\ref{hist_lam} is compatible with the combination of two sets of system, one with aligned orbits, and the other with randomly oriented orbits. As more systems are measured, most precise statistical properties of the distribution of orbital inclinations for close-in giant planets will become accessible.

An almost polar orbit of CoRoT-1b would be very suggestive of the action of an extreme dynamical event, such as a close encounter with another massive planet. Formation of close-in gas giants by migration in a gas disc would mainly result in aligned orbits \citep{lin96}. Gravitational interactions with other bodies  can perturb the spin-orbit alignment both during and after the planet formation stage, either through orbital angular momentum exchange (e.g. Kozai mechanism) or more dramatically through catastrophic resonances and close encounters. The first process is, for instance, a plausible explanation for the extremely eccentric system HD 80606. It is less likely to operate on CoRoT-1b, given its close and circular orbit and the absence of detected stellar or planetary perturber. The second type of process is a more likely explanation.

It has long been suggested that planet-planet scattering may account for the presence of close-in exoplanets \citep{chat08,ford08}. High spin-orbit angles would seem to favour this hypothesis, which has also gained renewed support from other lines of evidence:  multiple planetary systems tend to be dynamically dense \citep{raym09},
 the distribution of planetary eccentricities resembles that expected from a scattering process \citep{juri08}, 
 and the closest orbital distances  conform to the scattering scenario \citep{ford06}. The case of CoRoT-1 illustrates the high interest of vigorously pursuing the effort to measure the RM anomaly in transiting exoplanets.

\section*{Acknowledgments}

We thank Geoff Marcy for providing information on the optimal HIRES setup. We are grateful to the European Southern Observatory for allocation of time to this programme.  The Keck observations were made possible by a special allocation of 1 night to the CoRoT Key Science project. The authors wish to recognize the  cultural role and reverence that the summit of Mauna Kea has always had within the indigenous Hawaiian community.  We are most fortunate to have the opportunity to conduct observations from this mountain.

\bibliographystyle{mn2e}
\bibliography{corot1}{}

\end{document}